\UseRawInputEncoding
\documentclass[lettersize,correspondense]{IEEEtran}

\usepackage{amsmath,amsfonts}
\usepackage{array}
\usepackage[caption=false,font=normalsize,labelfont=sf,textfont=sf]{subfig}
\usepackage{textcomp}
\usepackage{stfloats}
\usepackage{url}
\usepackage{verbatim}
\usepackage{graphicx}
\usepackage{epstopdf}
\usepackage{amsmath}
\usepackage{algorithm}
\usepackage{algorithmicx}
\usepackage{algpseudocode}
\usepackage{float}

\usepackage{cite}

\usepackage[colorlinks=true,
            linkcolor=black, 
            anchorcolor=black, 
            citecolor=black, 
            urlcolor=blue
            ]{hyperref}

\def\BibTeX{{\rm B\kern-.05em{\sc i\kern-.025em b}\kern-.08em
    T\kern-.1667em\lower.7ex\hbox{E}\kern-.125emX}}
\usepackage{balance}

\begin{document}
\bibliographystyle{ieeetr}
\newcommand{\upcite}[1]{\textsuperscript{\cite{#1}}}

\title{\Large Hidden Backdoor Attack against Deep Learning-Based Wireless Signal Modulation Classifiers}

\author{Yunsong Huang, Weicheng Liu and Hui-Ming Wang \emph{Senior Member, IEEE}

\thanks{
Copyright (c) 2015 IEEE. Personal use of this material is permitted. However, permission to use this material for any other purposes must be obtained from the IEEE by sending a request to pubs-permissions@ieee.org.

The work was supported by the National Natural Science Foundation of China under Grants 62171364 and 61941118. (\emph{Corresponding author: Hui-Ming Wang})

The authors are with the School of Information and Communications Engineering, Xi'an Jiaotong University,
Xi'an 710049, China, and also with the Ministry of Education Key Laboratory for Intelligent Networks and Network Security, Xi'an Jiaotong
University, Xi'an 710049, China (e-mail: song1102@stu.xjtu.edu.cn; liuweicheng@stu.xjtu.edu.cn; xjbswhm@gmail.com).}}

\maketitle

\begin{abstract}
Recently, DL has been exploited in wireless communications such as modulation classification. However, due to the openness of wireless channel and unexplainability
of DL, it is also vulnerable to adversarial attacks. In this correspondence, we investigate a so called hidden backdoor attack to modulation classification, where the adversary puts elaborately  designed poisoned samples on the basis of IQ sequences into training dataset.
These poisoned samples are hidden because it could not be found by traditional classification methods.
And poisoned samples are same to samples with triggers which are patched samples in feature space.
We show that the hidden backdoor attack can reduce the accuracy of modulation classification significantly with patched samples. At last, we propose activation cluster to detect abnormal samples in training dataset.
\end{abstract}


\section{Introduction}
\IEEEPARstart{M}{odulation} classification is a critical issue in many civilian and military communications systems. In civil applications, it is mainly used in signal identification, spectrum management, electronic surveillance and interference confirmation for wireless spectrum management \cite{sills1999maximum}. In military applications, it is necessary to identify the modulation schemes of unknown signals when implementing radio jamming and deception \cite{peng2021survey}. Moreover, modulation classification is regarded as one of the key techniques for communications systems with adaptive modulation capabilities \cite{peng2021survey}.

Modulation classification methods are mainly divided into maximum likelihood (ML) hypothesis test based on decision theory and statistical pattern recognition based on feature extraction. Some researchers use ML decision theory to identify modulation in \cite{sills1999maximum},\cite{kim1988digital},\cite{swami2000hierarchical}. Deep learning (DL) has become an efficient method in extracting features. By capturing the intrinsic characteristics of the spectrum data, DL can be applied to raw signals and can effectively operate using feature learning and latent representations. In \cite{peng2021survey}, researchers convert the raw modulated signals into images that have a grid-like topology and use convolutional neural networks (CNN) to classify the modulation scheme. In \cite{2020DNN}, a recurrent neural network (RNN) classifier is introduced to combat the Rayleigh fading as well as uncertain noise and achieve robustness to the increase of frequency offset. In \cite{2019OFDM}, IQ sequences in Orthogonal Frequency Division Multiplexing (OFDM) systems are handled by a CNN  to achieve accurate, consistent, and robust modulation classification.


However, due to the openness of the wireless propagations and the black-box feature of the NN, there are several types of attack against the modulation classification. For example, evasion attack aims to fool the DL model into making wrong decisions in the test phase, where fast gradient sign method (FGSM) \cite{usama2019adversarial} and Carlini-Wagner algorithm (CWA) \cite{kokalj2019targeted} are proposed.
Generative adversarial network (GAN) can generate a spoofing signal which confuses the classification model and mislead the classification results \cite{davaslioglu2018generative}.
Poisoning attack reduces the accuracy of a model by adding poisoned samples into training dataset \cite{peng2021survey}. Trojan attack rotates constellation points and changes the original label, then put the changed signals in training dataset to finish attack \cite{trojan}.

Recently, a new kind of poisoning attack called backdoor attack shows its severe threat to DL. Backdoor attack differs from GAN and FGSM that it manipulates only a selected number of training samples with specific and controled triggers in the entire life cycle of the DL system. The trigger means a small perturbation onto the normal traning samples added by the adversary. Compared to a general poisoning attack, the backdoor attack is more hard to be detected. This is because the classifier performs well on samples without triggers but will output specific misleading results under samples with triggers.
The backdoor attack attracts great attention in computer vision. However, so far there is not enough attention on backdoor attack in wireless signal modulation classification.

In this correspondence, we go a step further to introduce a so called hidden backdoor attack in wireless signal modulation classification \cite{saha2020hidden}. The samples poisoned by hidden backdoor attack is even more covert and can not be detected by traditional methods, for example, higher order statistics (HOS).
In the training phase, poisoned samples are generated by combining triggers and normal samples in a special way. The goal of poisoning is to make one source sample with trigger visually the same as the target sample but the same as source sample added by trigger in feature space. In test phase, the classifier will provide misleading results on samples patched with triggers.
Finally we propose a contermeasure to hidden backdoor attack, where activation cluster is exploted to detect whether training dataset is under attack.

\section{DL Model for Classification}
\subsection{The Signal Model}
Before we propose the hidden backdoor attack in wireless signal modulation classification, we first provide the signal model and the DL classification model.
Consider a received signal with a complex envelope
\begin{equation}
y(k)=A e^{j\left(2 \pi f_{0} k T+\theta_{k}\right)} \cdot \sum_{l=-\infty}^{\infty} x(l) h\left(k T-l T+\epsilon_{T} T\right)+g(k),
\end{equation}
where $x(l)$ represents the symbol sequence, $l$ is used to express which one in symbol sequence, $A$ is an unknown amplitude factor, $f_{0}$ denotes the carrier frequency offset, $\theta_{k}$ is the phase jitter, $k$ is used to express which one in received IQ sequence, $T$ represents the symbol spacing, $h(\cdot)$ denotes the residual channel effect, $\epsilon_{T}$ is the time error, and $g(k)$ represents the additive Gaussian noise. The goal of modulation classification is to determine which modulation scheme has been used by $y(k)$ from a candidate modulation set.

According to the DL-based modulation classification mentioned before, the received signal $y(k)$ is first preprocessed for signal representation and then fed into DL classification models to infer the modulation scheme. In this correspondence, we use the received signals in IQ sequences to train the DL classification model. The IQ point can be expressed like:
\begin{equation}
\begin{aligned}
y(k) &=I(k)+j \cdot Q(k) = a(k) \cdot e^{j \phi(k)}
\end{aligned}
\end{equation}
where $I(k)$, $Q(k)$, $a(k)$, and $\phi(k)$ represent the instantaneous in-phase component, quadrature component, amplitude, and phase of $y(k)$, respectively. In the DL model, 128 signals are used to detect the modulation scheme, i.e., $\boldsymbol{y}=[y(1), \cdots, y(128)]$. After putting IQ sequences into the training dataset, the DL classification model will learn the IQ samples in feature space and classify.

\subsection{The VT-CNN2 DL Classification Model}
\begin{figure}[t]
	\centering
	\includegraphics[width=3.2 in]{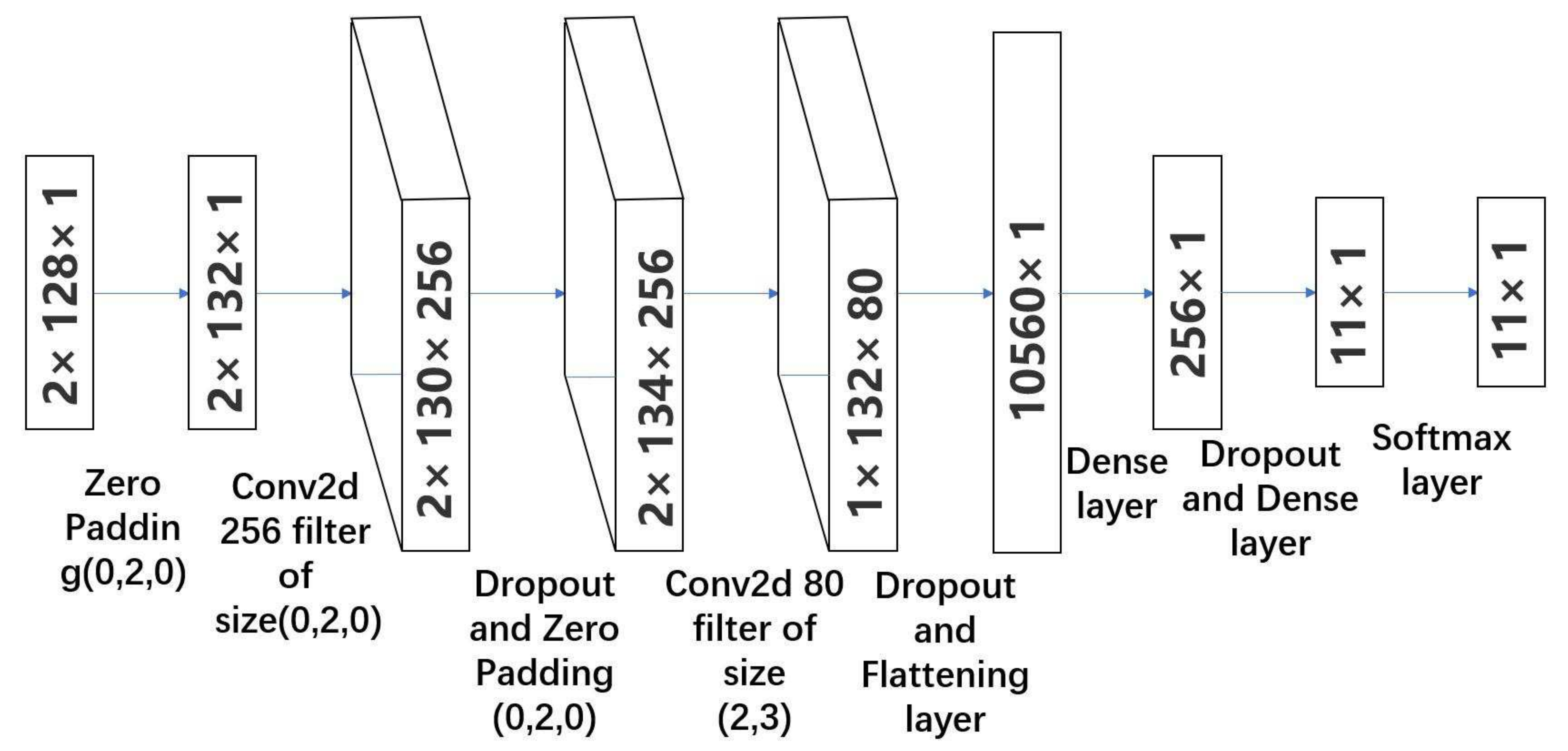}
	\caption{An illustration of VT-CNN2.}
	\label{fig1}
\end{figure}

To perform the modulation classification of the signal, we use a deep CNN classifier named as VT-CNN2 \cite{sadeghi2018adversarial}, which is illustrated in Fig. \ref{fig1}, following TensorFlow default format for data, i.e., height, width and channels.

There is a widely used IQ sequence dataset named RML2016.10A \cite{2016radio}. It consists of 220000 samples, where each sample is one kind of specific modulation at a specific signal to noise ratio (SNR). The dataset consists of BPSK, QPSK, 8PSK, 16QAM, 64QAM, BFSK, CPFSK, and PAM4 for digital modulations, and WB-FM, AM-SSB, and AM-DSB for analog modulations. The samples are generated for 20 different SNR from -20dB to 18dB with a step of 2dB. Each sample is a vector of size 2*128, we can expand this in IQ graph, corresponding to 128 in-phase and 128 quadrature components. The experiments are based on this dataset.














\section{The Attack Model}
For the purpose of showing the process of attack on VT-CNN2 DL classification model in details, we propose an attack scenario and the steps of achieving hidden backdoor attack.
We consider a scenario where the Tx (transmitter) sends signals to the Rx (receiver), and the adversary decides the time to send trigger to Rx.
Because Rx uses open dataset to train its DL model to classify the received signals, the dataset may be under the hidden hackdoor attack. If the DL model is under attack, it will provide specific wrong result when the Rx receives samples with triggers. But it still outputs accurate result when receiving normal samples.


In order to describle the attack clearly, in the following we take an example that the adversary wants to confuse a 8PSK constellation to a BPSK.
We define the 8PSK samples in the training set as source constellation $\mathcal S$
and each sample as source sample $\boldsymbol{y}_{\mathrm{source}}(k)\in \mathcal S, k=1,2,\cdots, K$, and define the BPSK samples in the training set as target constellation $\mathcal T$
and each sample as target sample $\boldsymbol{y}_{\mathrm{target}}(k)\in \mathcal T, k=1,2,\cdots, K$.
We call the samples generated by poisoning method as poisoned samples $\boldsymbol{y}_{\mathrm{p}}(k)$ and the source samples patched with triggers in random locations as patched source samples $\boldsymbol{y}^{*}(k)$.

\textbf{Step 1. Poisoning:}
We genarate a random vector $\boldsymbol{tr}$ of size $s$ where each element is generated obeying uniform random distribution. Different size of triggers will influence the performance of attack.
With $\boldsymbol{tr}$ we generate trigger  $\boldsymbol{t}$
which is in the form of
\begin{equation}
\boldsymbol{t}^T = \left[ {\bf 0}_{j},  \boldsymbol{tr}^T, {\bf 0}_{n-s-j} \right], \label{GT}
\end{equation}
where $n=128$ and $j$ means the location of $\boldsymbol{tr}$ in $\boldsymbol{t}$ which satisfies $ j \in [0,n-1-s]$ and $j $ is randomly generated. The trigger will be used to poison clean samples in training dataset.

With the trigger $\boldsymbol{t}$,
we generate poisoned sample $\boldsymbol{y}_{\mathrm{p}}(k)$ and patched source sample $\boldsymbol{y}^{*}(k)$ as, respectively:
\begin{equation}
\boldsymbol{y}^{*}(k) = \boldsymbol{y}_{\mathrm{source}}(k) + \boldsymbol{t}, \label{PS}
\end{equation}
\begin{equation}
\boldsymbol{y}_{\mathrm{p}}(k)=\mathcal{P}(\boldsymbol{y}^{*}(k), \boldsymbol{y}_{\mathrm{target}}(k)), \label{PA}
\end{equation}
where $\mathcal{P}$ is the poisoning algorithm.

The main idea of the algorithm  $\mathcal{P}(\cdot)$ is to put a perturbation on $\boldsymbol{y}_{\mathrm{target}}$  so that the obtained $\boldsymbol{y}_{\mathrm{p}}$ is similar to patched source samples $\boldsymbol{y}^{*}$ in feature space but similar to $\boldsymbol{y}_{\mathrm{target}}$ in IQ sequence. The similarities between $\boldsymbol{y}_{\mathrm{p}}$ and $\boldsymbol{y}^{*}$ in feature space make it possible to misclassify received samples $\boldsymbol{y}$. The detailed design of  $\mathcal{P}(\cdot)$ will be discussed in section IV.

\textbf{Step 2. Training:}
The adversary puts the poisoned samples $\boldsymbol{y}_{\mathrm{p}}$ to open  dataset which will be used by the Rx to train their classification model. After training with poisoned dataset, the DL model will misclassify patched 8PSK received samples to target BPSK. Meanwhile DL classification model performs well on normal samples. It is difficult to find DL model being under hidden backdoor attack only by the performance of DL model on normal samples.

\textbf{Step 3. Attacking:}
In the online test phase, Tx sends 8PSK sample $\boldsymbol{y}$ to the Rx.
Meanwhile, the adversary sends the patch $\boldsymbol{t}$ to the Rx which will be superpositioned to the 8PSK signal $\boldsymbol{y}$, such that the received signal at the Rx is $\boldsymbol{y}_{\mathrm{r}}^{*} $ in the form of
 \begin{equation}
 \boldsymbol{y}_{\mathrm{r}}^{*}=\boldsymbol{y} + \boldsymbol{t}, \label{PR}
 \end{equation}
where $\boldsymbol{y}_{\mathrm{r}}^{*}$ is defined as patched received sample. The goal of hidden backdoor attack is to mislead the classifier to make a decision that $\boldsymbol{y}_{\mathrm{r}}^{*}$  in (\ref{PR}) is a target BPSK sample.
Note that $\boldsymbol{y}_{\mathrm{r}}^{*}$ is different from $\boldsymbol{y}^{*}(k)$ in (\ref{PS}), $\boldsymbol{y}^{*}(k)$ is generated in training dataset and $\boldsymbol{y}_{\mathrm{r}}^{*}$ is generated by $\boldsymbol{y}$ which may be not the same as samples in training dataset.

The critical point of hidden backdoor attack is to design the poisoning algorithm  $\mathcal{P}(\cdot)$. We will discuss it in more details.

\section{Poisoning algorithm}

\subsection{Generating Poisoned Samples}

We assume that a pre-trained deep learning-based VT-CNN2 is used at a receiver to classify the received sample $\boldsymbol{y}$. We define DL model classifier $f_{\boldsymbol{\theta}}(.): \mathcal{Y} \rightarrow \mathcal{F}$, where $\boldsymbol{\theta}$ represents the model parameters, $\mathcal{Y} \subset \mathbb{C}^{n}$ is the received samples with $n$ being the dimension of the inputs, and $\mathcal{F} \subset \mathbb{R}^{m}$ indicates the feature vectors of $\mathcal{Y}$ and $m$ is the dimensions of $\mathcal{F}$. We call the feature vector generated by VT-CNN2 as $\boldsymbol{v}$. 

With DL model classifier $f_{\boldsymbol{\theta}}(.)$, the feature vector $\boldsymbol{v}$ of sample $\boldsymbol{y}$ can be described as $\boldsymbol{v}=f_{\boldsymbol{\theta}}(\boldsymbol{y})$.
We define $\boldsymbol{p}(k)$ which is a small perturbation on target sample $\boldsymbol{y}_{\mathrm{target}}(k)$. The goal of $\mathcal{P}(\cdot)$ is to generate a proper $\boldsymbol{p}(k)$ on $\boldsymbol{y}_{\mathrm{target}}(k)$ to make $f_{\boldsymbol{\theta}}(\boldsymbol{y}_{\mathrm{target}}(k) + \boldsymbol{p}(k))$ similar to $ f_{\boldsymbol{\theta}}(\boldsymbol{y}^{*}(k))$ and outputs $\boldsymbol{y}_{\mathrm{p}}(k)$ in the form of
\begin{equation}
 \boldsymbol{y}_{\mathrm{p}}(k) = \boldsymbol{y}_{\mathrm{target}}(k) + \boldsymbol{p}(k). \label{YP}
 \end{equation}

Define feature vectors as $\boldsymbol{v}_{\mathrm{1}}(k)=f_{\boldsymbol{\theta}}(\boldsymbol{y}^{*}(k))$ and $\boldsymbol{v}_{\mathrm{2}}(k)=f_{\boldsymbol{\theta}}(\boldsymbol{y}_{\mathrm{p}}(k))$, and
loss function is defined as $\mathcal{L}(\boldsymbol{v}_{\mathrm{1}}(k),\boldsymbol{v}_{\mathrm{2}}(k),\boldsymbol{p}(k))=\left\|\boldsymbol{v}_{1}(k)-\boldsymbol{v}_{2}(k)\right\|_{2}^{2}$ to measure the difference between $\boldsymbol{y}^{*}(k)$ and $\boldsymbol{y}_{\mathrm{p}}(k)$ in feature space.
Therefore, we formulate the optimization problem to generate the perturbation $\boldsymbol{p}(k)$ as the solution of
\begin{eqnarray}
\begin{split}
\min _{\boldsymbol{p}(k)} & \left\|\boldsymbol{v}_{1}(k)-\boldsymbol{v}_{2}(k)\right\|_{2}^{2}, \\
\text { s. t. } & \|\boldsymbol{p}_i(k)_{\mathrm{}}\| \leq \delta, \text{for all}\  i. \label{Co}
\end{split}
\end{eqnarray}
where the subscript $i$ means the $i$-th element of $\boldsymbol{p}(k)$. Since $f_{\boldsymbol{\theta}}(.)$ is not a convex function, so
the loss function $\mathcal{L}(\cdot)$ of CNN is not convex but a complex irregular function. It is very difficult to obtain the derivative of this function. Therefore, we perform one iteration of mini-batch projected gradient descent for the loss function and solve the problem as \cite{qian2015efficient}.
With such a small $\boldsymbol{p}(k)$, if we use traditional methods (e.g. higher order statistics) to classify $\boldsymbol{y}_{\mathrm{target}}(k)$ and $\boldsymbol{y}_{\mathrm{p}}(k)$, the two samples will be classified into one modulation scheme \cite{swami2000hierarchical}.
In such a way, $\boldsymbol{y}_{\mathrm{p}}(k)$ generated by $\mathcal{P}(\cdot)$ is similar to $\boldsymbol{y}_{\mathrm{target}}(k)$ in IQ sequence as (\ref{YP}) and similar to $\boldsymbol{y}^{*}(k)$ in feature space as (\ref{Co}).
We describe $\mathcal{P}(\cdot)$ in details in  Algorithm 1.

Because algorithm complexity varies with network complexity, we analyse complexity under our DL model. The complexity of Algorithm 1 is
\begin{eqnarray}
\begin{split}
\mathcal{O}\left(n\cdot\left(\sum_{l=1}^{C} M_{l} \cdot K_{l}\cdot c_{l i n}\cdot c_{l o u t}+\sum_{l=1}^{D} d_{l i n}\cdot d_{l o u t}\right)\right), \notag
\end{split}
\end{eqnarray}
where $n$ means the number of iterations, $M_{l}$ means the product of width and height of output characteristic graph in the $l_{th}$ convolution layer, $K_{l}$ means the product of width and height in $l_{th}$ convolution kernel, $c_{l in}$, $c_{l out}$ are the number of input channels and output channels respectively in $l_{th}$ convolution kernel. $d_{l in}$ and $d_{l out}$ are input size and number of output neurons in $l_{th}$ linear layer respectively. $C$ and $D$ are the number of convolution layers and dense layers respectively. And the poisoning process is deployed offline.
\begin{algorithm}[t]
        \caption{Generating Poisoned Samples}
        \begin{algorithmic}[1] 
            \State Input: learning rate $lr$, number of circles $num\_iter$, perturbation $\boldsymbol{p}(k)$, feature vector $\boldsymbol{v}_{\mathrm{1}}(k)$ and $\boldsymbol{v}_{\mathrm{2}}(k)$, maximum step size $\delta$, source sample $\boldsymbol{y}_{\mathrm{source}}(k)$, target sample $\boldsymbol{y}_{\mathrm{target}}(k)$;
            \State Create trigger $\boldsymbol{t}$ according to (\ref{GT});
            \State Patch $\boldsymbol{y}_{\mathrm{source}}(k)$ with $\boldsymbol{t}$ to get $\boldsymbol{y}^{*}(k)$ according to (\ref{PS});
            \For{$i$ in $num\_iter$}
              \State adjust $lr$
              \State $\boldsymbol{v}_{\mathrm{1}}(k)\gets f_{\boldsymbol{\theta}}(\boldsymbol{y}^{*}(k))$
              \State $\boldsymbol{v}_{\mathrm{2}}(k)\gets f_{\boldsymbol{\theta}}(\boldsymbol{y}_{\mathrm{target}}(k) + \boldsymbol{p}(k))$
              \State Find one-to-one mapping a(k) between $\boldsymbol{v}_{\mathrm{1}}(k)$ and $\boldsymbol{v}_{\mathrm{2}}(k)$
              \State $loss\gets \left\|\boldsymbol{v}_{1}(k)-\boldsymbol{v}_{2}(k)\right\|_{2}^{2}$
              \State Perform one iteration of mini-batch projected gradient descent for the loss function.
              \State $\boldsymbol{p}(k)\gets \boldsymbol{p}(k) - lr*\nabla_{\boldsymbol{p}(k)}$
              \If{$\|\boldsymbol{p}_i(k)_{\mathrm{}}\| \leq \delta$}
               \State $\boldsymbol{p}_i(k)_{\mathrm{}} \gets \boldsymbol{p}_i(k)_{\mathrm{}}$;
               \ElsIf{$\boldsymbol{p}_i(k)_{\mathrm{}} > \delta$ or $\boldsymbol{p}_i(k)_{\mathrm{}} < -\delta$}
               \State $\boldsymbol{p}_i(k)_{\mathrm{}} \gets \delta$ or $\boldsymbol{p}_i(k)_{\mathrm{}} \gets -\delta$;
              \EndIf
              \If{loss is small enough}
              \State break;
              \EndIf
           \EndFor
           \State $\boldsymbol{y}_{\mathrm{p}}(k)\gets \boldsymbol{p}(k) + \boldsymbol{y}_{\mathrm{target}}(k)$
        \end{algorithmic}
    \end{algorithm}

\subsection{The Results of Poisoned Samples}
We choose 8PSK as source class and BPSK as target class at the SNR = 18dB to show the generated poisoned samples. We set learning rate $lr = 0.00001$, number of cycles $num\_iter = 5000$ and maximum step size $\delta = 0.0001$. The trigger $tr$ obeys normal distribution with trigger size $s = 40$.

\begin{figure}[t]
\centering
\includegraphics[width=3.5in]{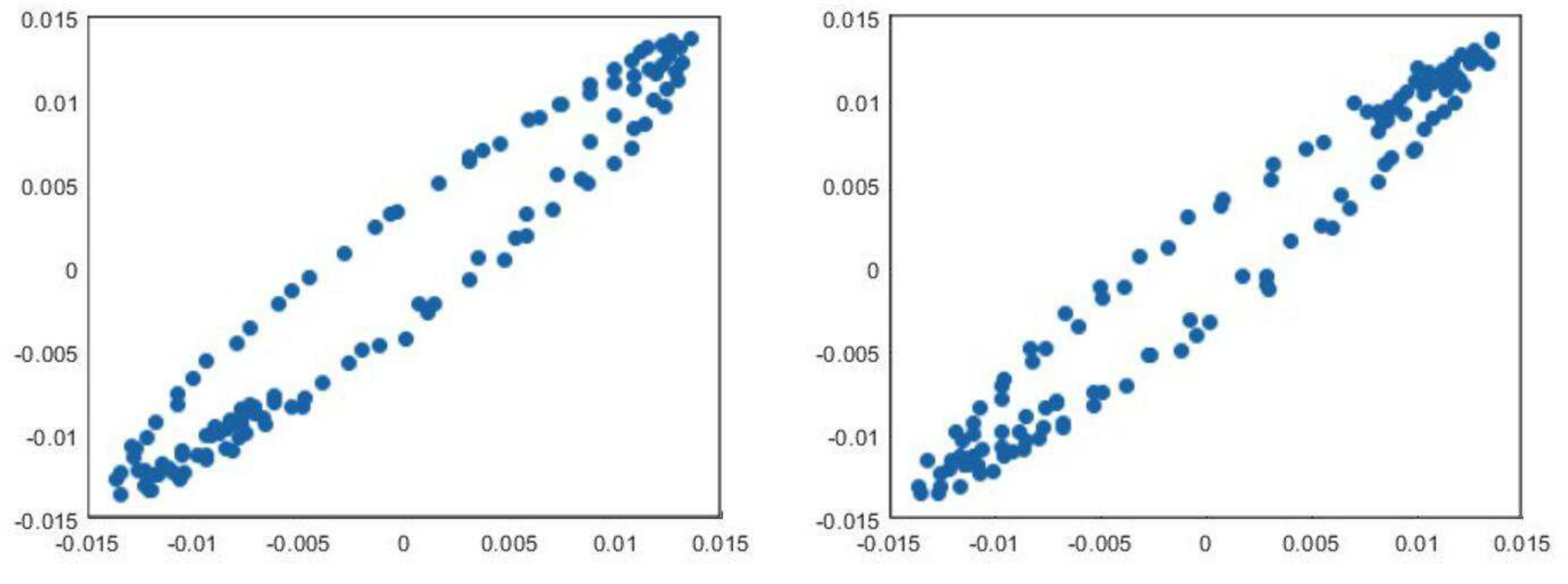}
\caption{Comparison between target sample (the left one) and poisoned sample (the right one) in constellation graph by hidden backdoor attack.}
\label{fig4}
\end{figure}

Fig. \ref{fig4} compares the constellation graph of target sample and poisoned BPSK sample generated by $\mathcal{P}(\cdot)$. The left one is a clean sample, the right one is a poisoned sample which is 8PSK in feature space but is similar to BPSK in constellation graph. If we use traditional methods (e.g.,HOS) to classify the samples, the classification results will show that these two samples are the same in modulation scheme \cite{swami2000hierarchical}.

Trojan attack also influences the DL model by poisoning training dataset, but they use the Givens rotation to generate poisoned samples \cite{trojan}. We define the generator matrix $\mathbf{G}_{\beta}$ as:
\begin{eqnarray}
\begin{split}
\mathbf{G}_{\beta}=
\left[\begin{array}{cc}
\cos \beta & \sin \beta \\
-\sin \beta & \cos \beta
\end{array}
\right],
\end{split}
\end{eqnarray}
where $\beta$ is the rotation angle. If input sample is $\boldsymbol{y}$, the poisoned sample can be describes as $\mathbf{G}_{\beta}\boldsymbol{y}$, then poisoned sample will be labeled as target. HOS has a natural robustness to constellation rotation and the fourth-order moments and cumulants do not change with the rotation \cite{swami2000hierarchical}. Using HOS to detect poisoned samples generated by simple rotation is particularly effective.
The more similar two samples are, the harder can the user detect the poisoned samples in target category. So we come to a conclusion that poisoned samples generated by $\mathcal{P}(\cdot)$ are difficult to be detected.


\subsection{The Possibility of Successful Attack}
In this section, we show numerical results on the performance of hidden backdoor attack.
We consider that poisoned dataset consists of 100 poisoned samples, 150 normal samples of target category and 250 normal samples of source category with each SNR. The percentage of poisoned samples of target category is 40\%.
We finetune a binary classifier for the source and target modulation schemes. After training with poisoned dataset, the adversary will send trigger to the Rx to launch the attack, and the Rx receives 200 patched received samples to test. We call the attack successful if the DL model classifies the patched samples into the wrong label. We call the classification failure if the DL model misclassifies the clean samples.

\begin{figure}[t]
\centering
\includegraphics[width=2.5 in]{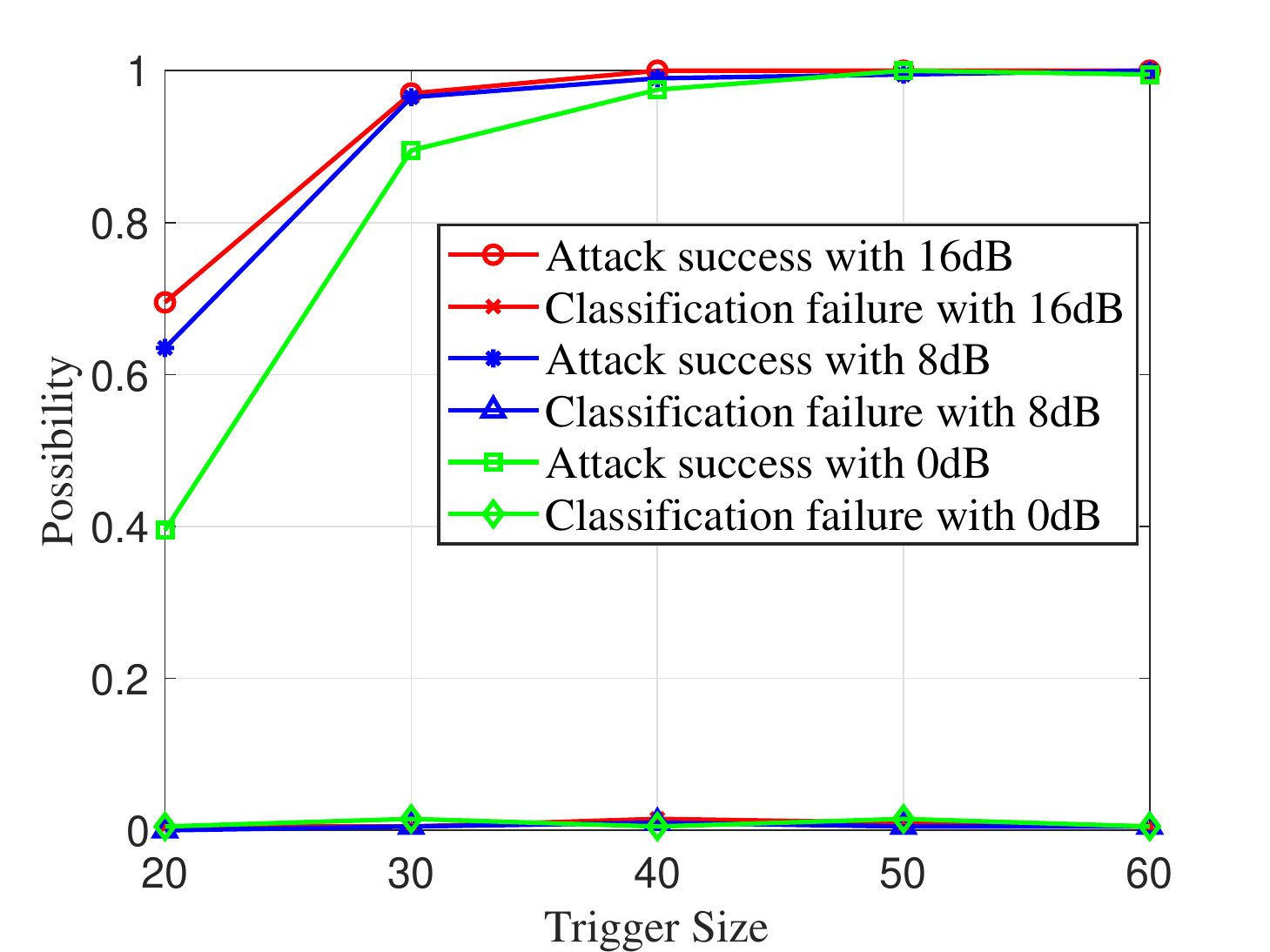}
\caption{Attack success and classification failure probability with poisoned samples in different SNR with different size of triggers.}
\label{fig6}
%
%
\centering
\includegraphics[width=2.5 in]{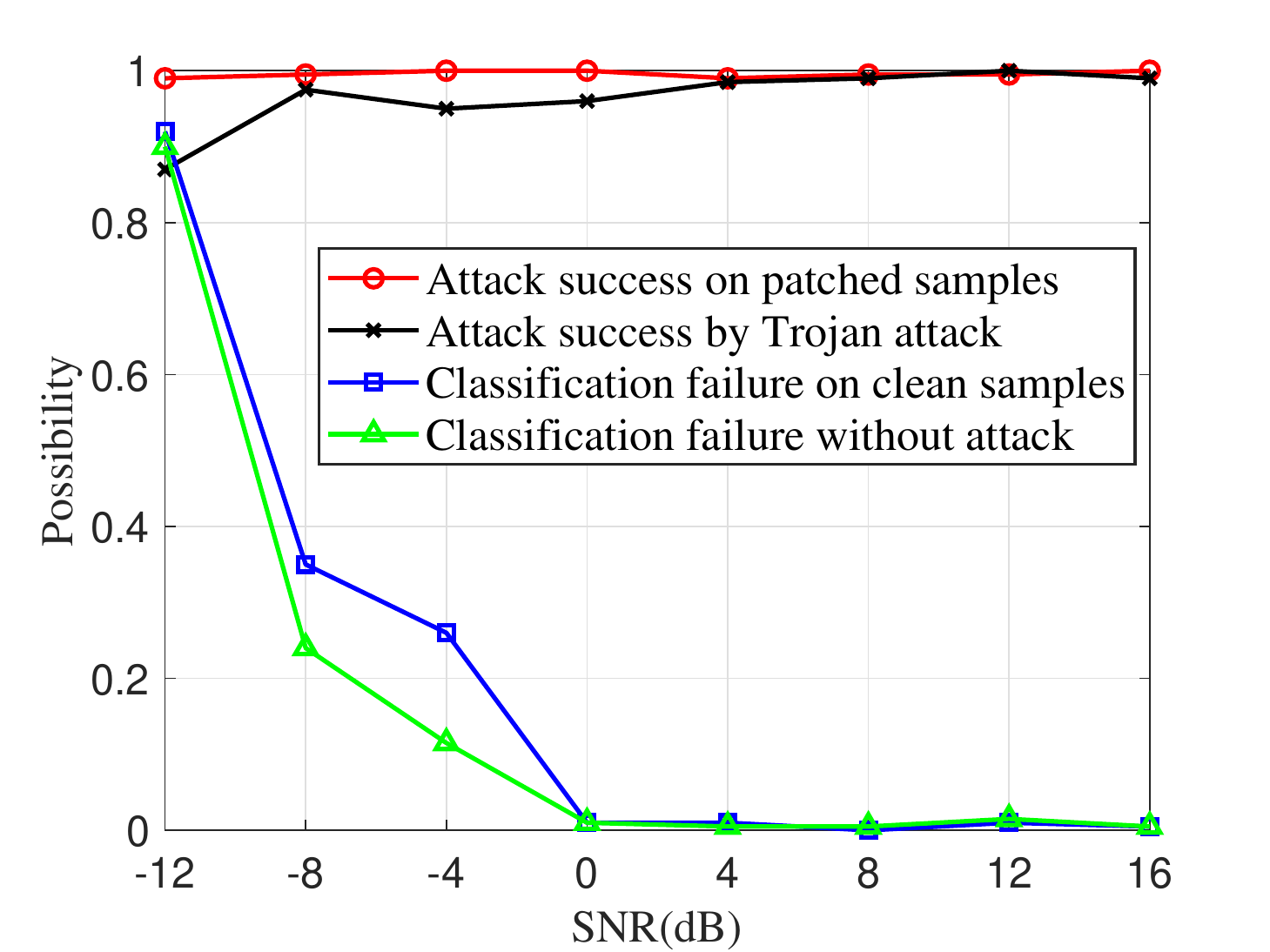}
\caption{Attack success probability and classification failure probability under hidden backdoor attack.}
\label{fig7}
\end{figure}

Fig. \ref{fig6} depicts the attack success and classification failure possibility versus size of trigger $\boldsymbol{tr}$. The results are averaged over 200 patched received samples. It is observed that the size of trigger will impact the attack success probability significantly. Because with larger size of trigger, the difference  between $\boldsymbol{y}_{\mathrm{p}}(k)$ and $\boldsymbol{y}_{\mathrm{target}}(k)$ in feature space can be easier to be identified, thus the attack success possibility will be higher. And the classification failure possibility stays relatively low with trigger size varying from 20 to 60. The results verify that attack success possibility is higher with larger size of trigger. So we set trigger size $s$ as 40 to confirm attack successful.

Fig. \ref{fig7} depicts the attack success and classification failure possibility versus SNR under setting $\boldsymbol{tr}=40$. The results are averaged over 200 patched received samples and 200 clean source samples. In addition, attack success possibility by Trojan attack is simulated under 400 poisoned samples as a comparison \cite{trojan}. We can see that the possibility of successful attack is almost 100\% all the time. Hidden backdoor attack can attack successfully on patched samples and SNR does not have much influence on the performance of hidden backdoor attack with trigger size $s=40$. Meanwhile, DL model classification failure possibility is low on clean samples with SNR more than $-5$dB. Consider that a device seldom receives signals with SNR being lower than 0dB, it is hard to detect that the trained model has been poisoned only by the classification results on clean samples.

\section{Detecting Hidden Backdoor Attack}
Since traditional detection method which is successful in detecting Trojan attack such as HOS could not detect the hidden backdoor attack, we now propose an effective countermeasure to detect the poisonded sample in the hidden backdoor attack.

\subsection{The Activation Cluster}
In this section, we propose activation cluster to detect poisoned samples in training dataset.
For clean samples that are originally target classes, the model identifies the features learned from the input samples of target classes. For patched received samples, the DL network identifies the characteristics related to the poisoned samples generated by $\mathcal{P}(\cdot)$, which leads to the misclassification of patched received samples into the target class by model.
The idea of activation cluster is to use the difference in mechanism which can be verified in the activation of the DL network, and the activation of the last hidden layer of the model can be clustered to distinguish \cite{chen2018detecting}.

We consider that the DL model $f_{\boldsymbol{\theta}_p}: \mathcal{Y} \rightarrow \mathcal{F}$ is trained with untrusted training dataset $D_{p}$, the network is queried using the training data and the resulting activations of the last hidden layer are retained. The resulting activations of the last hidden layer can be used to classify the modulation scheme. And analyzing the activations of the last hidden layer is enough to detect poison \cite{chen2018detecting}. So we use the last hidden layer to detect the poisoned samples.

We define $\boldsymbol{A}(k)$ which holds activation for all $\boldsymbol{y}(k)$ such that the DL model will classify $\boldsymbol{y}(k)$ into target class. $\boldsymbol{y}_{\mathrm{p}}(k)$ and $\boldsymbol{y}_{\mathrm{target}}(k)$ are similar in constellation graph but different in feature space, so $\boldsymbol{y}_{\mathrm{p}}(k)$ and $\boldsymbol{y}_{\mathrm{target}}(k)$ will be clustered into two clusters.
To cluster the activations, we first reshape the activations $\boldsymbol{A}(k)$ and perform dimensionality reduction using independent component analysis (ICA). Then we perform K-means with $k = 2$ to separate the poisoned from legitimate activations \cite{chen2018detecting}. If training samples in one modulation scheme can be clustered into two clusters, we should consider that there are poisoned samples in training dataset.


\subsection{The Detection Results}
\begin{figure}[t]
\centering
\includegraphics[width=3in]{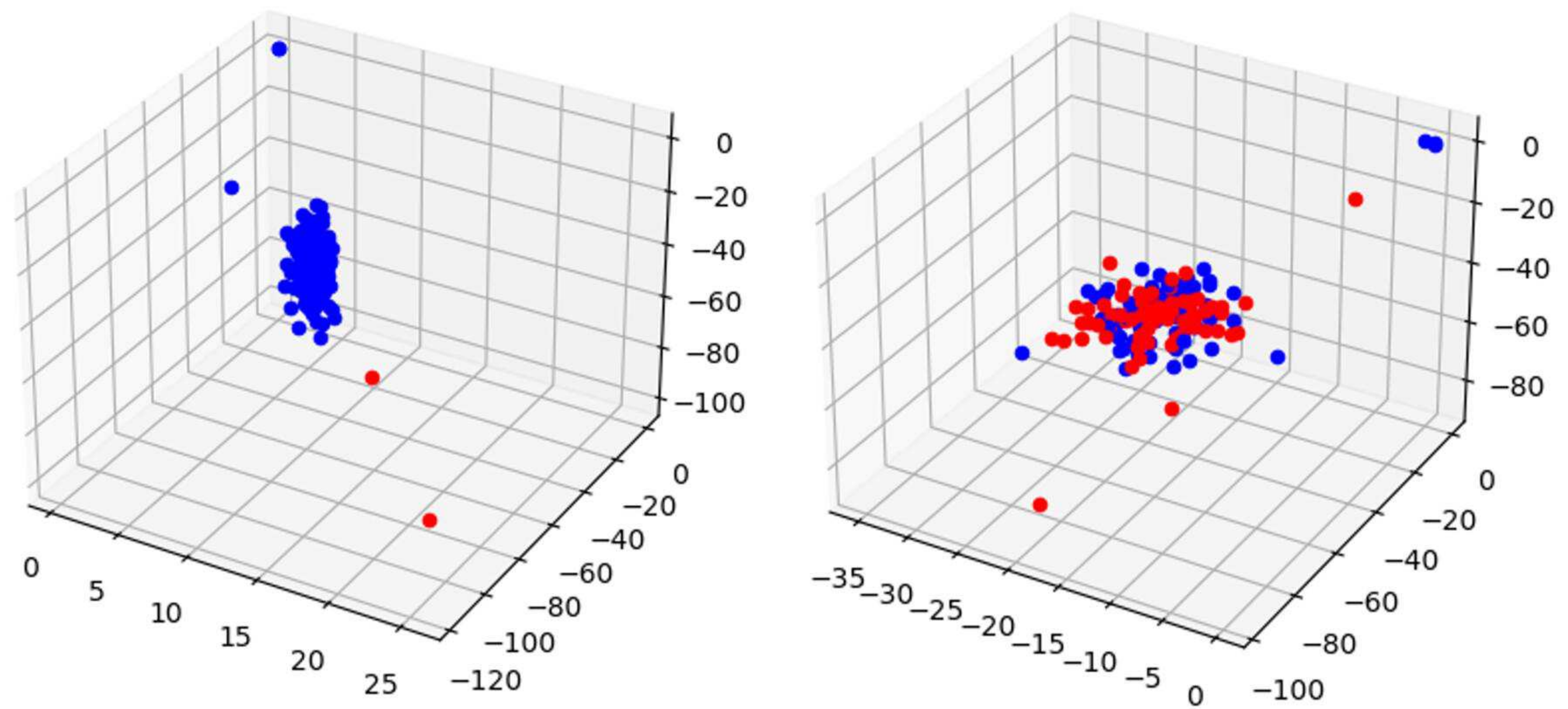}
\caption{The results of the activation cluster on poisoned dataset. (The labels are source in left and target in right)}
\label{fig10}
\centering
\includegraphics[width=2.5in]{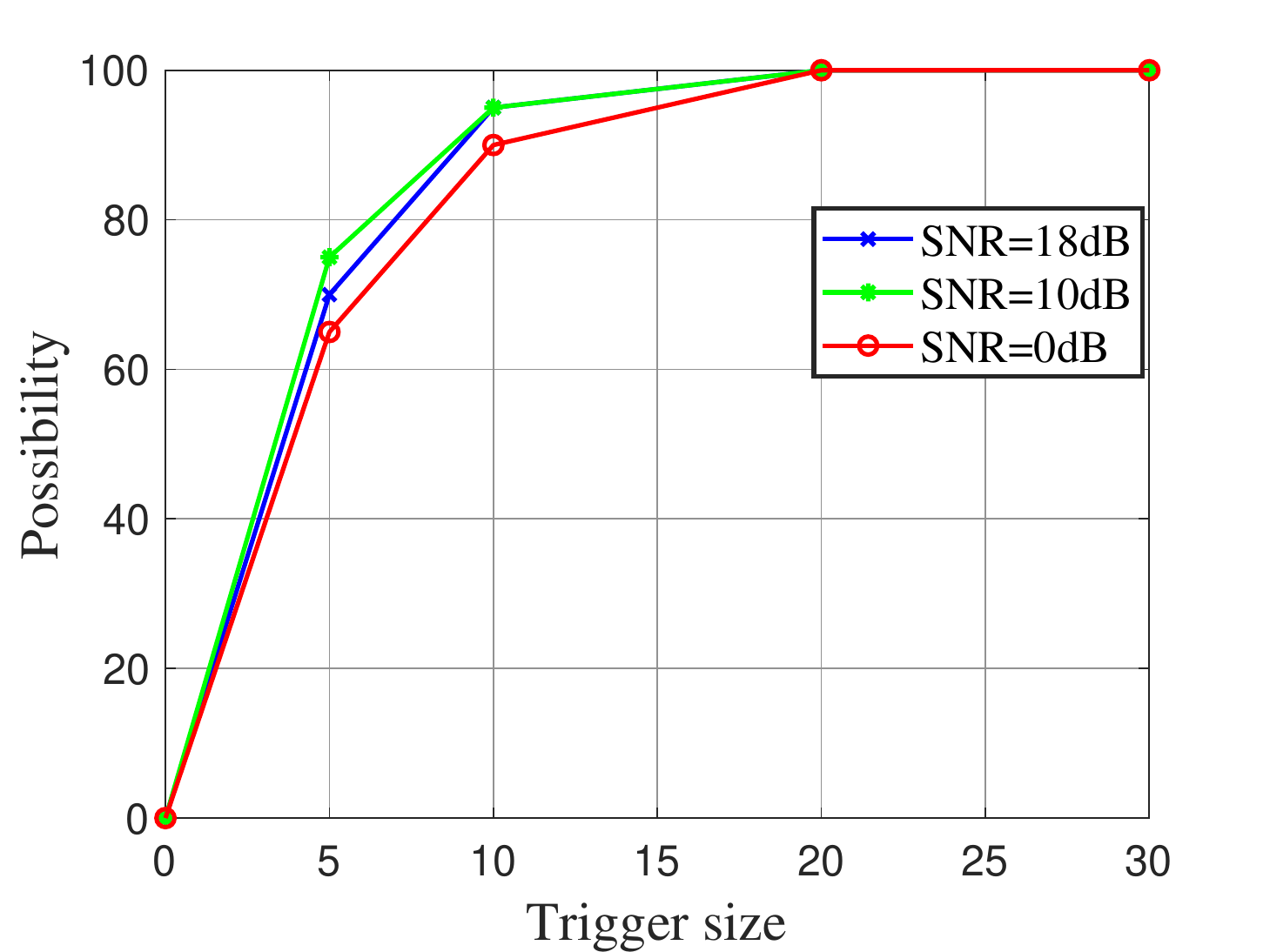}
\caption{The success possibility of activation cluster on poisoned dataset.}
\label{fig11}
\end{figure}

Fig. \ref{fig10} shows the cluster results on 150 source samples, 150 target samples and 100 poisoned samples which have a target label with SNR$\ =18$dB. We can observe that samples are divided into one cluster in source label, but samples are divided into two clusters in target label. That means that there are poisoned samples in dataset.

Fig. \ref{fig11} depicts the activation cluster success possibility versus trigger size $\boldsymbol{tr}$ on poisoned dataset. The results are averaged over 20 times under different SNR. When size of $\boldsymbol{tr}$ increasing, the detection success possibility increases, for more obvious difference in feature space between poisoned samples and clean targeted samples. But this method is not sensitive to SNR. Because triggers are generated independently with the SNR of samples and we assume that triggers are not influenced by channel information.
We have already simulated the attack success probability versus the size of triggers in Fig. \ref{fig6}. In order to ensure the successful attack, the size of $\boldsymbol{tr}$ is usually set more than 25. So the activation cluster can detect poisoned dataset generated by hidden backdoor attack reliably.

\section{Conclusion}
This correspondence introduced hidden backdoor attack against DL-based wireless signal modulation classifiers. We compared this method with Trojan attack and results show that hidden backdoor attack is better in avoiding being detected.
After training with poisoned dataset, the classification result of DL model is right when receiving normal samples and wrong when receiving patched samples. Simulation results have shown that attack success possibility could achieve almost 100\% under different SNR with a relatively big trigger size. In addition, we proposed activation cluster to detect the hidden backdoor attack by using the resulting activations of the last hidden layer. The cluster results showed that activation cluster can reliably detect poisoned samples in dataset.

Based on this work, there are some research directions that can be further explored. First, the hidden backdoor attack can be used to attack other classification applications based on DL model, such as RF (radio frequency) fingerprint identification. Second, other effective defence countermeasures could be developed against the hidden backdoor attack.




\begin{thebibliography}{1}\balance

\bibitem{sills1999maximum} J. A. Sills, ``Maximum-likelihood modulation classification for PSK/QAM," in \emph{Proc. IEEE MILCOM}, vol. 1, pp. 217-220, 1999.

\bibitem{peng2021survey} S. Peng, S. Sun, and Y.-D. Yao, ``A Survey of Modulation Classification Using Deep Learning: Signal Representation and Data Preprocessing," \emph{IEEE Trans. Neural Netw. Learn. Syst.}, early access, Jun.14 2021, doi:{\color{blue}
\href{http://dx.doi.org/10.1109/TNNLS.2021.3085433}{10.1109/TNNLS.2021.3085433}}.

\bibitem {kim1988digital} K. Kim and A. Polydoros, ``Digital modulation classification: the BPSK versus QPSK case," in \emph{Proc. IEEE MILCOM}, pp. 431-436, 1988.






\bibitem {swami2000hierarchical} A. Swami and B. M. Sadler, ``Hierarchical digital modulation classification using cumulants," \emph{IEEE Trans. on Commun.}, vol. 48, no. 3, pp. 416-429, 2000.


\bibitem {2020DNN} S. Hu, Y. Pei, P. P. Liang, and Y.-C. Liang, ``Deep neural network for robust modulation classification under uncertain noise conditions," \emph{IEEE Trans. Veh. Technol.}, vol. 69, pp. 564-577, Jan. 2020.

\bibitem {2019OFDM} S. Hong \emph{et al.}, ``“Deep Learning-Based Signal Modulation Identification in OFDM Systems," \emph{IEEE Access}, vol. 7, pp. 114631-114638, 2019.

\bibitem {usama2019adversarial} M. Usama, M. Asim, J. Qadir, A. Al-Fuqaha, and M. A. Imran, ``Adversarial machine learning attack on modulation classification," in \emph{Proc. U.K./China Emerg. Technol. (UCET)}, Oct. 2019, pp. 1-4.

\bibitem {kokalj2019targeted} S. Kokalj-Filipovic, R. Miller, and J. Morman, ``Targeted adversarial examples against RF deep classifiers," in \emph{Proc. ACM Workshop Wireless Secur. Mach. Learn. (WiseML)}, 2019, pp. 6-11.



\bibitem {davaslioglu2018generative} Z. Wang, W. Liu and H. -M. Wang, "GAN Against Adversarial Attacks in Radio Signal Classification," in \emph{IEEE Commun. Lett.}, vol. 26, no. 12, pp. 2851-2854, Dec. 2022, doi:{\color{blue}
\href{http://dx.doi.org/10.1109/LCOMM.2022.3206115}{10.1109/LCOMM.2022.3206115}}.




\bibitem {trojan} K. Davaslioglu and Y. E. Sagduyu, ``Trojan attacks on wireless signal classification with adversarial machine learning," in \emph{Proc. IEEE Int. Symp. Dyn. Spectr. Access Netw. (DySPAN)}, 2019, pp. 1-6.

\bibitem {saha2020hidden} A. Saha, A. Subramanya, and H. Pirsiavash, ``Hidden trigger backdoor attacks," in \emph{Proc. AAAI}, 2020, pp. 11957-11965.

\bibitem {sadeghi2018adversarial} M. Sadeghi and E. G. Larsson, ``Adversarial attacks on deep-learning based radio signal classification," \emph{IEEE Wireless Commun. Lett.}, vol. 8, pp. 213-216, Feb. 2019.

\bibitem {2016radio} T. J. O’shea and N. West, ``Radio machine learning dataset generation with gnu radio," in \emph{Proc. GNU Radio Conf.}, 2016, vol. 1, no. 1.

\bibitem {qian2015efficient} Q. Qian, R. Jin, J. Yi, L. Zhang, and S. Zhu, ``Efficient distance metric learning by adaptive sampling and mini-batch stochastic gradient descent (SGD)," \emph{Mach. Learn.}, vol. 99, no. 3, pp. 353-372, 2015.

\bibitem {chen2018detecting} B. Chen \emph{et al.}, ``“Detecting backdoor attacks on deep neural networks by activation clustering," in \emph{Proc. AAAI Workshop}, 2019.



\end{thebibliography}


\end{document}